\documentclass[12pt,onecolumn]{IEEEtran}
\usepackage{graphicx}
\usepackage{ amsmath }
\usepackage{ amssymb }
\usepackage{amsfonts}
\usepackage{amscd}
\usepackage{ latexsym }
\newtheorem{thm}{Theorem}[section]
\newtheorem{prop}[thm]{Proposition}

\newtheorem{example}[thm]{Example}
\newtheorem{theorem}{Theorem}[section]
\newtheorem{proposition}[thm]{Proposition}

\newtheorem{note}[thm]{Note}

\newtheorem{remark}[thm]{Remark}

\title{Quantum Quasi-Cyclic LDPC Codes}
\author{
\authorblockN{Manabu Hagiwara\authorrefmark{1}\authorrefmark{2} and Hideki Imai\authorrefmark{1}\authorrefmark{3}}\\
\authorblockA{\authorrefmark{1}National Institute of Advanced Industrial Science and Technology,\\
Research Center for Information Security,\\
Akihabara-Daibiru Room 1102,\\
1-18-13 Sotokanda, Chiyoda-ku, Tokyo, Japan,\\
Email: hagiwara.hagiwara@aist.go.jp, h-imai@aist.go.jp.}\\
\authorblockA{\authorrefmark{2}Center for Research and Development Initiative,\\
Chuo University,\\
1-13-27 Kasuga, Bunkyo-ku, Tokyo, Japan.}\\
\authorblockA{\authorrefmark{3}
Department of Electrical, Electronic and Communication Engineering, 
Faculty of Science and Engineering, Chuo University,\\
1-13-27 Kasuga, Bunkyo-ku, Tokyo, Japan.}\\}
\begin{document}
\maketitle
\sloppy

\begin{abstract}
In this paper, a construction of a pair of ``regular'' quasi-cyclic LDPC codes to construct a quantum error-correcting code is proposed.
In other words, we find quantum regular LDPC codes with various weight distributions.
Our construction method is based on algebraic combinatorics and achieves a lower bound of the code length, and has lots of variations for length, code rate.
These codes are obtained by a descrete mathematical characterization for model matrices of quasi-cyclic LDPc codes.
\end{abstract}


\section{Introduction}
Quantum error-correcting codes provide detection or correction of the errors which occur in a communication through a noisy quantum channel.
Hence the codes protect integrity of a message sent from a sender to a receiver.
This paper presents a construction method of a pair of ``regular'' quasi-cyclic low-density parity-check codes (regular QC-LDPC codes) as ingredients of a CSS code.
QC-LDPC codes are known as a practical class of classical error-correcting codes due to its compact representation and good performance, especially for short code lengths \cite{Tanner,hagi:Fossorier}.
Since CSS codes find their applications not only for quantum error-correction but also for privacy amplification of quantum cryptography, they have become important research objects for quantum information theory \cite{shor-preskill}.

Mackay proposed bicycle codes which are constructed by a combination of heuristic method and theoretical approach \cite{hagi:mackay}.
In his paper, we find the following requirement ``we delete rows using the heuristic that the column weight of a matrix should be as uniform as possible''.
This implies that the design of weight distribution for quantum LDPC codes by deterministic method is a theoretically interesting problem.
Recently, various construction methods of quantum LDPC codes have been proposed \cite{hagi:tillich,hagi:tan,hagi:sal1,hagi:sal2,hagi:hsieh,hagi:camara}.
Among these proposed constructions, it has been difficult to design the weight distribution of the related parity-check matrices.
In this paper, we give a solution for the weight distribution design problem for regular weight case.

Contributions of this paper are the following:
1. We find a characterization for model matrices of QC-LDPC codes with circulant permutation matrices to be ingredient codes of a CSS code.
The characterization is easily treatable for not only theoretical use but also computer experiments.
2. We propose a construction for $(\lambda_1, \rho)$-regular and $(\lambda_2, \rho)$-regular parity-check matrices for any $1 \le \lambda_1, \lambda_2 \le \rho/2$ such that the associated QC-LDPC codes are ingredient codes of a CSS code.
In other words, we find quantum regular LDPC codes.
Our method for designing a pair of LDPC matrices is theoretical and deterministic.
Furthermore, our construction achieves a theoretical bound on the length for quasi-cyclic LDPC code to have girth at least 6.
We show that our proposed codes satisfy an optimality of a bound which is arose from a difference matrix theory and has been investigated in classical LDPC code theory.
3. Computer experiences show outstanding error-correcting performance for our proposed codes.
In fact, the performances of our proposed codes almost achieve a bounded distance decoding bound, which is thought as a theoretical limit for algebraic codes, with various quantum code rates.


\section{Classical Quasi-Cyclic LDPC Codes}
Throughout this paper, we assume that a classical linear code is defined over $\mathbb{F}_2 := \{ 0,1 \}$, i.e. the code is a binary code.
In this section, we introduce a quasi-cyclic LDPC code with circulant permutation matrices.

\subsection{Classical Quasi-Cyclic LDPC Codes with Circulant Permutation Matrices}
Let $P$ be a positive integer.
Let $I(\infty)$ be a zero matrix of size $P$ and $I(1)=(a_{j, l})$ a matrix of size $P$ such that $a_{j, l} = 1$ if $l-j = 1$ and $a_{j, l}=0$ otherwise:
\[ I(1) = 
\left(
  \begin{array}{ccccc}
       & 1   &    &    &    \\
       &    &  1  &    &    \\
       &    &    & \ddots   &    \\
       &    &    &    &  1  \\
    1   &    &    &    &    \\
  \end{array}
\right).\]
For an integer $b$, put $I(b) := I(1)^{b}$.
The matrix $I(b)$ is called a \textbf{circulant permutation matrix}.
The integer $b$ is called the \textbf{index} of a circulant matrix $I(b)$.

A linear code $C$ is called a \textbf{quasi-cyclic (QC) LDPC code} (with circulant permutation matrices),
if a parity-check matrix $H_{C}$ of $C$ has the following block form:
\[
H_{C} =
\left[
  \begin{array}{cccc}
    I(c_{0,0})    & I(c_{0,1})     & \dots   & I(c_{0, L-1})   \\
    I(c_{1,0})    & I(c_{1,1})     & \dots   & I(c_{1, L-1})   \\
    \vdots        & \vdots         & \ddots  & \vdots          \\
    I(c_{J-1, 0}) & I(c_{J-1,1})   & \dots   & I(c_{J-1, L-1})   \\
  \end{array}
\right],
\]
where $c _{j,l} \in [P_{\infty}] := \{0,1, \dots, P-1\} \cup \{\infty \}$.
We call such a matrix $H_{C}$ a \textbf{QC-LDPC matrix}.
As it is known, a parity-check matrix is not determined uniquely for a given $C$.
On the other hand, we would like to characterize a QC-LDPC code by a parity-check matrix. Therefore we describe a QC-LDPC codes as a pair $(C, H_C)$.
Furthermore, we regard that $(C, H_C)$ and $(C, H_C')$ are different LDPC codes for different QC-LDPC matrices $H_C$ and $H_C'$ even if they have the same code space $C$.

Let $\mathcal{H}_C$ denote a matrix which consists of the indices of $H_C$,
in other words,
\[
\mathcal{H}_{C} =
\left[
  \begin{array}{cccc}
    c_{0,0}    & c_{0,1}     & \dots   & c_{0, L-1}   \\
    c_{1,0}    & c_{1,1}     & \dots   & c_{1, L-1}   \\
    \vdots     & \vdots      & \ddots  & \vdots          \\
    c_{J-1, 0} & c_{J-1,1}   & \dots   & c_{J-1, L-1}   \\
  \end{array}
\right].
\]
We call $\mathcal{H}_{C}$ the \textbf{model matrix} of $H_{C}$.
It should be noted that a model matrix $\mathcal{H}_C$ characterizes a parity-check matrix $H_C$ of a quasi-cyclic LDPC code.

\begin{note}
Generally speaking, it is known that QC-LDPC codes have fruitful advantages for LDPC code theory.
One of advantages is the memory size for storing the parity-check matrix.
Widely meaning, an \textbf{LDPC code} $(C, H_C)$ is defined as a kernel space $C$ associated with a low-density matrix $H_C$, where a low-density matrix is a matrix such that that most of elements are zero.
It is possible to construct a low-density parity-check matrix randomly.
Imagine a randomly constructed low-density parity-check matrix.
Because of the randomness, large size memory is required to store the parity-check matrix.
On the other hand, it is not the case for QC-LDPC codes because the parity-check matrix is reconstructed by its model matrix.
\end{note}

\begin{note}
(Classical) QC-LDPC codes have advantage not only from the viewpoint of memory, but also from the viewpoint of error-correcting performance \cite{hagi:ieee802}.
In particular, with sum-product decoding, it is shown that the error-correcting performances of short length QC-LDPC codes, e.g. of length 100, 1,000, 10,000, are similar to the ones of random LDPC codes with the same lengths \cite{hagi:Fossorier}.
In practical use, the available length of error-correcting code depends on a communication system.
In fact, short length QC-LDPC codes are chosen for real communication systems, e.g. WiMAX and DVB-s2 \cite{hagi:ieee802,hagi:DVB-s2}.
\end{note}

Our main goal of this paper is to propose a quantum error-correcting codes based on classical quasi-cyclic LDPC codes.
By nowadays technologies, it is impossible to implement quantum error-correcting codes.
For (very) future, we hope quantum error-correcting codes are used for practical use.
In such an age, it is meaningful to have a table of quantum error-correcting codes with various lengths.
Therefore it is interesting to construct quantum error-correcting codes based on quasi-cyclic LDPC codes.

\subsection{Regular LDPC Codes}\label{hagi:regularcase}
An LDPC code $(C, H_C)$ is called \textbf{$(\lambda, \rho)$-regular} (or \textbf{regular}, in short) if the numbers of 1's in any columns and any rows of a parity-check matrix $H_C= (  h_{j,l} )$ are constants $\lambda$ and $\rho$, respectively.
Formally, $\# \{ h_{j,l}=1 | 0 \le j < J \} = \lambda$ for any $0 \le l < L$ and $\# \{ h_{j,l}=1 | 0 \le l < L \} = \rho$ for any $0 \le j < J$, where $L$ (resp. $J$) is the number of columns (resp. rows).

The parameter $\lambda$ is called the \textbf{column weight} and $\rho$ is called the \textbf{row weight}.
At the beginning of the study of LDPC codes, the construction research has been focused on regular LDPC codes \cite{hagi:wada_ex1,hagi:wada_ex2,hagi:wada_ex3,hagi:wada_ex4,hagi:wada_ex5}.
For the design of regular LDPC codes, the following three have been regarded as important parameters:
the column weight $\lambda$, the row weight $\rho$, and \textbf{the girth}.
The definition of a girth is given in the next subsection.

\begin{example}\label{hagi:expl_regular}
Let $H_C$ be a binary matrix as follows:
\[
H_{C} =
\left[
  \begin{array}{cccccc}
    1    & 1     & 0   & 0 & 1 & 1   \\
    0    & 0     & 1   & 1 & 1 & 1   \\
    1    & 1    & 1   & 1 & 0 & 0   \\
  \end{array}
\right].
\]
The LDPC code $(C, H_C)$ is a $(2, 4)$-regular code.

Let $H_C'$ be a binary matrix as follows:
\[
H_C' =
\left[
  \begin{array}{cccccc}
    1    & 1     & 0   & 0 & 1 & 1   \\
    0    & 0     & 1   & 1 & 1 & 1   \\
  \end{array}
\right].
\]
Although the code space $C$ of $H_C'$ is the same as the code space of $H_C$,
the LDPC code $(C, H_C')$ is not a regular code.
\end{example}

\begin{proposition}
Let $C$ be a QC-LDPC code with a model matrix $\mathcal{H}_C$ and an LDPC matrix $H_C$.
If the symbol $\infty$ does not appear in the model matrix as an entry,
then the QC-LDPC code $(C, H_C)$ is a regular LDPC code.
In particular, the QC-LDPC code $(C, H_C)$ is a $(J, L)$-regular, where $J$ (resp. $L$) is the number of rows (resp. columns) of $\mathcal{H}_C$.
\end{proposition}
\begin{proof}
The LDPC matrix $H_C$ consists of $J$ row-blocks and $L$ column-blocks and each block is a circulant permutation matrix, i.e. non-zero matrix by the assumption.
A circulant matrix has unique 1 in each row and each column.
Therefore, $H_C$ has just $J$ 1s in each column and $L$ 1s in each row.
\end{proof}

\subsection{Girth of a Tanner graph}
The girth is one of properties for a Tanner graph of a binary matrix $(h_{j,l})$.
Therefore we would like to define Tanner graph here.
Formally speaking, a \textbf{Tanner graph} is defined as a pair of vertex sets associated with the indices of the rows $\{ 0, 1, \dots, J-1 \}$ and the columns $\{ 0, 1, \dots, L-1 \}$, and its subset $E$ as the edge set:
$$E := \{  (j, l) \in \{ 0, 1, \dots, J-1 \} \times \{ 0, 1, \dots, L-1 \} | h_{j, l} = 1 \}.$$

The girth of a Tanner graph is the smallest length of a cycle of the Tanner graph.
Since a Tanner graph is bipartite, the size of a cycle in the graph must be even and greater than or equal to 4.
If the girth of a Tanner graph for an LDPC matrix is 4, it tends that the error-correcting performance of the LDPC code is not outstanding. 
If the Tanner graph has no cycle, i.e. the graph is a tree, then it is expected that the performance of the \textbf{sum-product decoding}, which is a standard decoding algorithm for LDPC codes, is the same as that of a maximally likelihood decoding.
If a parity-check matrix is associated with a $(\lambda, \rho)$-regular LDPC code for $\lambda, \rho \ge 2$, a Tanner graph of the matrix cannot be a tree, in other words, it is unavoidable to contain cycles in the Tanner graph.
On the other hand, if there is a column or row whose Hamming weight is one, the sum-product decoding does not work (see the definition of sum-product decoding).
Therefore it is important to attain a big girth in the Tanner graph as one of the research directions (See Refs.~\cite{hagi:Hu} and \cite{Tanner} for details).

\begin{remark}\label{remark:girth_dual_containing}
Let $C$ be a \textbf{dual containing code}, i.e. $C^{\perp} \subset C$, where $C^{\perp}$ is the dual code of $C$. (The explicit definition of dual code is given in \ref{2-A}.)
It is known that the Tanner graph associated with any parity-check matrix of dual containing codes has a cycle of size 4.
Hence the girth is 4.
\end{remark}

\begin{example}
A parity-check matrix $H_C$ in Example \ref{hagi:expl_regular} has twelve 1s as entries.
Therefore, there are just 12 elements in the edge set $E$ of Tanner graph:
$E = \{ (0, 0), (0, 1), (0, 4), ( 0, 5), (1, 2), (1, 3), (1, 4), (1, 5), (2, 0), (2, 1),\linebreak[1] (2, 2), (2, 3) \}$.

A linear code $C$ associated with a parity-check matrix $H_C$ in Example \ref{hagi:expl_regular} is a dual containing code.
As it is mentioned in Remark \ref{remark:girth_dual_containing},there is a cycle of size 4 in the Tanner graph.
In fact, $(0,4), (1,4), (1, 5), (0,5)$ is a cycle of length 4.
\end{example}

In order to construct a code with good error-correcting performance, the girth should not be $4$.
Now we state the following condition and denote by (G):
\begin{itemize}\label{hagi:girth}
\item[(G)] The girth of the Tanner graph of a parity-check matrix is greater than or equal to 6.
\end{itemize}

\begin{proposition}[\cite{hagi:Fossorier}]\label{hagi:fact:marc}
A necessary and sufficient condition to satisfy (G) for a QC-LDPC code with the model matrix $\mathcal{H}_C = ( c_{j,l} )_{i,j}$ is
$ c_{j_{1},l_{1}} - c_{j_{1}, l_{2}} + c_{j_{2}, l_{2}} - c_{j_{2}, l_{1} } \neq 0 \pmod{P} $ for any $ 0 \le j_{1} < j_{2} < J, 0 \le l_{1} < l_{2} < L$, where $P$ is the size of the circulant permutation matrices.
\end{proposition}

\section{CSS Codes}\label{hagi:sec:CSS}
\textbf{CSS} (Calderbank-Shor-Steane) codes are quantum codes and are determined by a pair of classical linear codes $C$ and $D$ which satisfy the condition (T) below.
We start with introducing the condition (T) and classical quasi-cyclic LDPC codes.

\subsection{Twisted Condition (T) and CSS codes}\label{2-A}
Let $C$ and $D$ be classical linear codes.
Let us recall that the (classical) dual code of a linear code.
For a linear code $C$, the dual code $C^{\perp}$ is defined by
\[C^{\perp} = \{ x \in \mathbb{F}_{2}^n | x \times c^{T} = 0, \forall c \in C\}.\]
In other words, the \textbf{dual code} is a linear code generated by a parity-check matrix of the code: any codeword of the dual code is a linear combination of rows of the parity-check matrix.
For the linear codes $C$ and $D$, it is said that $C$ and $D$ satisfy the \textbf{twisted condition} (T) if  
$$ D^{\perp} \subset C, $$
or equivalently,
$$ C^{\perp} \subset D. $$
It is easy to verify that the condition (T) is equivalent to $H_C \times H_D^{T} = 0$, where $H_C$ and $H_D$ are parity-check matrices of $C$ and $D$ respectively.

Let $\mathbb{C}$ be a complex number field.
For a pair of classical linear codes $C$ and $D$ which satisfy (T), a CSS code \cite{hagi:CSs,hagi:csS} is defined as a Hilbert subspace $Q$ of $(\mathbb{C}^{2})^{\otimes n}$ spanned by 
$$ \sum_{d \in D^{\perp} } | c + d \rangle \in (\mathbb{C}^{2})^{\otimes n}, $$
for $c \in C$,
where $(\mathbb{C}^{2})^{\otimes n}$ is a tensor space of degree $n$ of $\mathbb{C}^{2}$, i.e. $( \mathbb{C}^{2} )^{\otimes n} = \overbrace{ \mathbb{C}^2 \otimes \mathbb{C}^2 \otimes \dots \otimes \mathbb{C}^2}^{n}$, $ \{ |0 \rangle, | 1 \rangle \} $ forms a computational basis of $\mathbb{C}^{2}$, and $| c_1 c_2 \dots c_n \rangle := |c_1 \rangle \otimes |c_2 \rangle \otimes \dots | c_n \rangle $ for $c_i \in \mathbb{F}_2 $.
The linear codes $C$ and $D$ are called \textbf{ingredient codes} of a CSS code $Q$.
Since the representatives of a coset $C / D^{\perp}$, as abelian groups, form a basis of the CSS code, the dimension of the CSS code $Q$ as a complex vector space is equal to $2^{\dim C - \dim D^{\perp} }$.

\subsection{Necessary and Sufficient Condition to Satisfy (T) for classical QC-LDPC Codes}
We give a combinatorial necessary and sufficient condition to satisfy the twisted condition (T) in terms of model matrices $\mathcal{H}_C$ and $\mathcal{H}_D$ of QC-LDPC codes $(C, H_C)$ and $(D, H_D)$ respectively.
Let us denote $j$th (resp. $k$th) rows of model matrices $\mathcal{H}_{C} = (c_{j,l})$ (resp. $\mathcal{H}_{D} = (d_{k,l}) $) of $H_C$ (resp. $H_D$) by $c_{j}$ (resp. $d_{k}$) in other words, $c_{j} := (c_{j,0}, c_{j,1}, \dots, c_{j, L-1} )$ and $d_{k} := (d_{k,0}, d_{k,1}, \dots, d_{k, L-1} ).$
For Example \ref{hagi:expl1}, $c_{0}=(1, 2, 4, 3, 6, 5)$ and $d_{1}=(1, 4, 2, 5, 6, 3).$
The necessary and sufficient condition is represented by a notion of ``multiplicity even'' defined as follows: for an integer sequence $x=(x_{0}, x_{1}, \dots, x_{L-1})$, 
we call $x$ \textbf{multiplicity even} if each entry appears even times in $x_0, x_1, \dots, x_{L-1}$ except for the symbol $\infty$.
For example, $(0, 1, 1, 0, 3, 3, 3, 3, \infty)$ is multiplicity even, but $(0,2,2,4,4,4,0)$ is not.

Before introducing the condition, let define the minus operation ``$-$'' to $[P_{\infty}]$ as follows:
for $x, y \in \{0, 1, \dots, P-1\}$, put 
\[x-y := x-y \pmod P,\]
 and
\[ x-\infty = \infty -x = \infty - \infty := \infty.\]
We naturally extend the operation $-$ to a sequence of $[P_{\infty}]$ by position-wise operation, in other words, we define the \textbf{minus operation} ``$-$'' between rows of model matrices
\begin{prop}\label{prop:trans_2_minus}
For any $x, y \in [ P_{\infty} ]$,
$I(x) I(y)^{T} = I(x-y)$.
\end{prop}
\begin{proof}
If $x = \infty$ or $y = \infty$, $I(x) I(y)^{T} = 0 = I(\infty) = I(x-y)$.
If $x, y \neq \infty$, $I(x) I(y)^{T} = I(x) I(-y) = I(1)^{x} I(1)^{-y} = I(1)^{x-y} = I(x-y)$.
\end{proof}
\begin{remark}
In \cite{hagi:ieee802}, the symbol ``-1'' is used for describing a zero matrix in stead of $\infty$.
As it is shown in Proposition \ref{prop:trans_2_minus}, it is more natural to use $\infty$ for zero-matrix when we operate circulant matrices.
Therefore we adapt $\infty$ in this paper.
\end{remark}

\begin{theorem}\label{hagi:iff:twisted}
Let $(C, H_C) $ and $(D, H_D)$ be QC-LDPC codes with model matrices $\mathcal{H}_{C}$ and $\mathcal{H}_{D}$ respectively such that they have the same circulant sizes.
The codes $(C, H_C)$ and $(D, H_D)$ satisfy the twisted condition if and only if
$c_{j} - d_{k}$ is multiplicity even for any row $c_{j}$ of $\mathcal{H}_{C}$ and any row $d_{k}$ of $\mathcal{H}_D$.
\end{theorem}
\begin{proof}
We divide $H_{C}$ into $J$ row-blocks $H_{C_{0}}, H_{C_{1}}, \dots, H_{C_{(J-1)}}$:
\[H_{C_{j}} := (I(c_{j,0}), I(c_{j,1}), \dots, I(c_{j, L-1}) ), 0 \le j < J,\]
where $J$ is the number of rows of the model matrix $\mathcal{H}_C$.
Similarly, we divide $H_{D}$ into $K$ row-blocks, where $K$ is the number of rows of the model matrix $\mathcal{H}_D$.

These codes $(C, H_C)$ and $(D, H_D)$ satisfy the twisted condition if and only if $H_{C} \times  H_{D} ^{\mathrm{T}} = 0$ as it is mentioned in the last paragraph of \ref{2-A}.
Clearly, $H_C \times H_D^T = 0$ is equivalent to $c \times d^{T} = 0 $ for any row $c$ of $H_{C_j}$ and any row $d$ of $H_{D_k}$ for any $0 \le j < J, 0 \le k < K$.

Denote $a$th row of $H_{C_{j}}$ by $\mathcal{C}_{a}^{(j)}$ and $b$th row of $H_{D_{k}}$ by $\mathcal{D}_{b}^{(k)}$, for $0 \le a,b < P$.
Remark that $\mathcal{C}_{a}^{(j)}$ and $\mathcal{D}_b^{(k)}$ are binary vectors.
Define a binary matrix $X_{j,k} = (x_{a, b})_{0 \le a, b, < P}$  by putting $X_{j, k} := H_{C j} \times  (H_{D k})^{\mathrm{T}}$.
By the construction of $X_{j, k}$, we have $x_{a, b} := \mathcal{C}_{a}^{(j)} \times  \mathcal{D}_{b}^{(k) \mathrm{T}}.$
By this notation, $X_{j, k}=0$ for $0 \le j < J$ and $0 \le k < K$ if and only if any row $\mathcal{C}_{a}^{(j)}$ of $H_{C_{j}}$ and any row $\mathcal{D}_{b}$ of $H_{D_{k}}$ are orthogonal to each other.

By Proposition \ref{prop:trans_2_minus}, we have $H_{C j} \times  H_{D k}^{\mathrm{T}} = \sum_{0 \le l < L} I(c_{j, l} - d_{k, l}).$
Since $I(x)$ is a binary circulant permutation matrix for any integer $x$,
$X_{j,k}=0$ if and only if $c_{j} - d_{k}$ is multiplicity even.
\end{proof}

Let $\mathcal{H}_C$ and $\mathcal{H}_D$ be model matrices as zero matrices.
Then the associated QC-LDPC codes $(C, H_C)$ and $(D, H_D)$ satisfy (T) by Theorem \ref{hagi:iff:twisted}.
Thus it seems easy to construct ingredient codes by a similar way.
Unfortunately, the associated quantum codes does not have a good error-correcting performance.

To obtain good performance codes, we require (T) and (G) simultaneously for ingredient codes.
In the next subsection, we give a necessary and sufficient condition to satisfy (G) for quasi-cyclic LDPC codes in terms of model matrices with minus operation.

\subsection{Characterization for (G) by the Minus Operation}
A term ``\textbf{multiplicity free}'' means all the entries of a given vector are different.
For example, $(0,1,2,3,4,5,6,7)$ is multiplicity free, however, $(0,0,1,2,3,4,5)$ is not.
We state the following:
\begin{proposition}\label{hagi:prop:free}
A QC-LDPC code $(C, H_C)$ with the model matrix $\mathcal{H}_C$ satisfy (G) if and only if 
$c_{j_{1}} - c_{j_{2}}$ is multiplicity free for any $0 \le j_{1} < j_{2} <J$, where $c_{j}$ is $j$th row of $\mathcal{H}_C$.
We omit the proof since it is directly obtained by Proposition \ref{hagi:fact:marc}.
\end{proposition}

By Theorem \ref{hagi:iff:twisted} and Proposition \ref{hagi:prop:free}, it is easy to verify two QC-LDPC codes satisfy (T) and (G) simultaneously from their model matrices.
\begin{example}
Put model matrices $\mathcal{H}_C$ and $\mathcal{H}_D$ with circulant matrices of size $4p$ by:
\[
\mathcal{H}_{C} = 
\left(
  \begin{array}{cccccc}
    0   & 0   & 0   & 0   \\
    0   & p   & 2p  & 3p   \\
  \end{array}
\right)
\mathcal{H}_{D}
=
\left(
  \begin{array}{cccccc}
    0   & 0   & 2p   & 2p  \\
    0   & p   & 0    & p  \\
  \end{array}
\right).
\]

Then the associated QC-LDPC codes $(C, H_C)$ and $(D, H_D)$ satisfy (T) and (G) simultaneously by Theorem \ref{hagi:iff:twisted} and Proposition \ref{hagi:prop:free}.
\end{example}

\section{Four-Cycle Codes}

\subsection{Four-Cycle Codes}
We call a matrix $T$ a \textbf{tire} if the matrix is a \textbf{circulant matrix} over $[P_{\infty}]$ (of size $L/2$):
\[
T = 
\left(
  \begin{array}{cccc}
  t_0     & t_1   & \dots   & t_{L/2-1}   \\
  t_{L-1} & t_0   & \dots   & t_{L/2-2}   \\
  \vdots     &          &         & \vdots   \\
  t_1     & t_2   & \dots   & t_0   \\
  \end{array}
\right),
\]
where $t_{i} \in [P_{\infty}]$.

Let $Q$ be a CSS code with ingredient codes $(C, H_C)$ and $(D, H_D)$ which are QC-LDPC codes.
We call $Q$ a \textbf{four-cycle code} if the model matrices $\mathcal{H}_C$ and $\mathcal{H}_D$ of $H_C$ and $H_D$, respectively, have the forms:
$$ \mathcal{H}_{C} = [ T_{1}, T_{2} ], \mathcal{H}_{D} = [ T_{3}, T_{4} ],$$
by some circulant matrices $T_{1}, T_{2}, T_{3}$ and $T_{4}$ over $[P_{\infty}]$.
Note that a circulant matrix is not assumed to be a permutation matrix.
In other words, all $t_0, t_1, \dots$ may be non-zero elements.

\begin{proposition}\label{hagi:prop:two_rs_mats}
Let $T_{A}$ and $T_{B}$ be tires over $[P_{\infty}]$ of size $L/2$.
Define model matrices $\mathcal{H}_{C}$ and $\mathcal{H}_{D}$ by putting:
$\mathcal{H}_{C} := [ T_{A}, T_{B}]$ and $\mathcal{H}_{D} := [-T_{B}^{\mathrm{T}} , -T_{A}^{\mathrm{T}} ].$

Let $(C, H_C)$ and $(D, H_D)$ be QC-LDPC codes associated with model matrices $\mathcal{H}_C$ and $\mathcal{H}_D$, respectively.
Then $(C, H_C)$ and $(D, H_D)$ satisfy the twisted condition (T).

In other words, we obtain a four-cycle code with ingredient codes $(C, H_C)$ and $(D, H_D)$.
\end{proposition}
\begin{proof}
Denote the $j$th row (resp. $k$th row) of $\mathcal{H}_C$ (resp. $\mathcal{H}_D$) by $c_j$ (resp. $d_k$).
Then we can write $c_{j} - d_{k} = (a_{-j} +b_{k}, a_{1-j} + b_{k-1}, \dots, a_{L/2-j-1}+ b_{k-L+1}, b_{-j} + a_{k}, b_{1-j}+a_{k-1}, \dots, b_{L/2-j-1} + a_{k-L/2+1})$, where
$(a_{0}, a_{1}, \dots, a_{L/2-1})$ and $(b_{0}, b_{1}, \dots, b_{L/2-1})$ are the 0th rows of $T_{A}$ and $T_{B}$ respectively.

By arranging the order of entries, the entries of the left-half of $c_{j} - d_{k}$ are 
\[\{a_{0}+b_{k-j}, a_{1}+b_{k-j-1}, \dots, a_{L/2-1}+b_{k-j-L/2+1} \}, \]
 and
the entries of the right-half of $c_{k-1} - d_{j-1}$ are 
\[\{a_{0} + b_{k-j}, a_{1} + b_{k-j-1}, \dots, a_{L/2-1}+b_{k-j-L/2+1} \}.\]
Hence $c_{j} - d_{k}$ is multiplicity even.

By theorem \ref{hagi:iff:twisted}, LDPC codes $(C, H_C)$ and $(D, H_D)$ satisfy (T)
if and only if 
$c - d$ is a multiplicity even vector for any row $c$ of the model matrix $\mathcal{H}_C$ and a row $d$ of the model matrix $\mathcal{H}_D$.
\end{proof}

LDPC codes $(C, H_C)$ and $(D, H_D)$ are called \textbf{equivalent} (up to position-permutation) if $H_C$ is obtained by column and row permutations to $H_D$.
In other words, there exists a permutation matrix $P_R$ and $P_C$ such that $H_C = P_R H_D P_C$.

\begin{proposition}\label{hagi:two_structure}
The linear codes $C$ and $D$ in Proposition \ref{hagi:prop:two_rs_mats} are equivalent up to permutation.
\end{proposition}
\begin{proof}
Put a matrix $P_R := ( r_{i,j} )_{0 \le i, j < P L/2}$ and $P_C:=( p_{i,j})_{ 0 \le i, j < PL} $ by:
$$r_{i,j} = \delta_{i,PL/2 - j},  p_{i,j} = \delta_{i, PL - j},$$
where $\delta_{x, y}$ is the Kronecker's delta, i.e. $\delta_{x,y} = 1 $ for $x=y$ and $\delta_{x, y} = 0$ for $x \neq y$.
Then it is easy to verity that $H_C = P_R H_D P_C$.
\end{proof}

\subsection{Bicycle Code (MacKay's Code)}
The study of quantum LDPC codes was firstly developed by MacKay et.al.
In his paper \cite{hagi:mackay}, they proposed the following construction for ingredient LDPC codes:
\[
H_C = H_D =
\left[
  \begin{array}{cc}
   A    &  A^{\mathrm{T}}  \\
  \end{array}
\right],
\]
where
$H_C$ and $H_D$ are parity-check matrices of linear codes $C$ and $D$ respectively, $A$ is a binary circulant matrix,
 in other words, $A$ is a sum of circulant permutation matrices.
The CSS code is  called a \textbf{bicycle code}.

\begin{prop}
A bicycle code is a four-cycle code.
\end{prop}
\begin{proof}
Let $T_A$ be a tire over $[1_{\infty} ]$, i.e. $[P_{\infty}]$ with $P=1$.
Then $[1_{\infty}] = \{ 0, \infty \}$ as a set.
Define a model matrix $\mathcal{H}_C := [ T_A \; \; T_A^{\mathrm{T}} ]$ and $\mathcal{H}_D := [ -T_A \; \; -T_{A}^{\mathrm{T}} ]$.
We have $\mathcal{H}_D = [ -(T_A^{\mathrm{T}})^{\mathrm{T}} \;\; -T_A^{\mathrm{T}} ] = [ T_A \;\;  T_A^{\mathrm{T} } ] = \mathcal{H}_C$, since $T_A = -T_A$ and $( T_A^{ \mathrm{T} } )^{ \mathrm{T} } = T_A $.
\end{proof}

\section{Systematic Construction for Ingredient Codes with Girth More Than or Equal to 6}
\subsection{Mathematical Preparation}
The integers module $P$, denoted $\mathbb{Z}_P$, is a set of (equivalence classes of) integers $\{0, 1, \dots, P-1\}$.
Addition, minus operation, and multiplication in $\mathbb{Z}_P$ are performed module $P$ for elements of $\mathbb{Z}_P$.
From here, we denote the \textbf{greatest common divisor} for integers $a, b$ by $\mathrm{gcd}(a,b)$.

We quote the following fundamental facts from group theory and number theory:

\begin{proposition}[\cite{hagi:group}]\label{hagi:quotient}
Let $G$ be a group and $H$ a subgroup of $G$.
Put $[ g ] := \{ h g | h \in H\}$ for $g \in G$.
Then the followings are equivalent:\\
$$1) \;\; [g] \cap [g'] \neq \emptyset,$$
$$2) \;\; [g] = [g'],$$
$$3) \;\; g g'^{-1} \in H,$$
$$4) \;\; g' g^{-1} \in H.$$
\end{proposition}

\begin{proposition}[\cite{hagi:group}]\label{hagi:gp1}
For any positive integer $P \ge 2$, 
$\mathbb{Z}_{P}^{*} := \{ z \in \mathbb{Z}_{P} | \mathrm{gcd}(z, P) = 1 \}$ is an abelian (i.e. commutative) group with multiple operation of $\mathbb{Z}_{P}$.
In particular, for any $z \in \mathbb{Z}_P^*$, there exists $y \in \mathbb{Z}_{P}^*$ such that $yz = 1 \pmod P$.
\end{proposition}

\begin{proposition}[\cite{hagi:group}]\label{hagi:cyclic}
If $P$ is a prime, then $\mathbb{Z}_{P}^{*}$ is a cyclic group.
In other words, there exists a generator $z$ of $\mathbb{Z}_{P}^{*}$ such that 
$\mathbb{Z}_{P}^{*} = \{ 1, z, z^2, \dots z^{P-2} \} = \{ 1, 2, \dots, P-1 \}$.
\end{proposition}

\begin{theorem}[Dirichlet's Theorem \cite{hagi:dirichlet}]\label{hagi:gauss}
Let $a$ and $b$ be positive integers such that $\mathrm{gcd}(a, b)=1$.
Let $A_{n} := b + a  n$ for $n \ge 1$.
Then there are infinitely many primes in the series $A_{1}, A_{2}, \dots$.
\end{theorem}

Let $P, \sigma, \tau$ be positive integers.
For an integer $x$, $\sigma$ is said to be \textbf{$x$-affine coprime to $P$} if $x - \sigma$ (or equivalently $\sigma - x$) is coprime to $P$, i.e. $\mathrm{gcd}( x - \sigma, P ) = 1$.
We call an integer $\sigma$ a \textbf{fulfillment} to $P$ if $\sigma$ is coprime to $P$ and and $\sigma^{i}$ is 1-affine coprime to $P$ for $ 1 \le i < \mathrm{ord}_P(\sigma)$ where $\mathrm{ord}_P(\sigma)$ is the order of $\sigma$ as an element of $\mathbb{Z}_{P}^{*}$.
For example, $\sigma := P-1$ is a fulfillment to $P$ for $P \ge 2$ with $\mathrm{ord}_P (\sigma) = 2$.
Tables \ref{hagi:tbl-fullfill-1}, \ref{hagi:tbl-fullfill-2}, \ref{hagi:tbl-fullfill-3} and \ref{hagi:tbl-fullfill-4} are lists of the fulfillments $\sigma$ to $P$ under the condition $3 \le \mathrm{ord}(\sigma) \le 20$ and $ \sigma < P < 200$.
It will be useful to construct an ingredient pair obtained from Theorem \ref{hagi:main1}.

\begin{table}[htb]
\begin{center}
\begin{tabular}{l|c|r}
\hline\hline
\multicolumn{1}{c|}{\bf $\mathrm{ord}_P (\sigma)$} & {\bf $ P $}& {\bf $\sigma$}\\
\hline
3 & 7 & 2, 4 \\
\cline{2-3}
  & 13 & 3, 9 \\
\cline{2-3}
  & 19 & 7, 11 \\
\cline{2-3}
  & 31 & 5, 25 \\
\cline{2-3}
  & 37 & 10, 26 \\
\cline{2-3}
  & 43 & 6, 36 \\
\cline{2-3}
  & 49 & 18, 30 \\
\cline{2-3}
  & 61 & 13, 47 \\
\cline{2-3}
  & 73 &  8, 64 \\
\cline{2-3}
  & 79 & 23, 55 \\
\cline{2-3}
  & 91 & 9, 16, 74, 81 \\
\cline{2-3}
  & 97 & 35, 61 \\
\cline{2-3}
  & 103 & 46, 56 \\
\cline{2-3}
  & 109 & 45, 63 \\
\cline{2-3}
  & 127 & 19, 107 \\
\cline{2-3}
  & 133 & 11, 30, 102, 121 \\
\cline{2-3}
  & 139 & 42, 96 \\
\cline{2-3}
  & 151 & 32, 118 \\
\cline{2-3}
  & 157 & 12, 144 \\
\cline{2-3}
  & 163 & 58, 104 \\
\cline{2-3}
  & 169 & 22, 146 \\
\cline{2-3}
  & 181 & 48, 132 \\
\cline{2-3}
  & 193 & 84, 108 \\
\cline{2-3}
  & 199 & 92, 106 \\
\hline
4 & 13  & 5, 8 \\
\cline{2-3}
  & 17 &  4, 13 \\
\cline{2-3}
  & 25 &  7, 18 \\
\cline{2-3}
  & 29 & 12, 17 \\
\cline{2-3}
  & 37 &  6, 31 \\
\cline{2-3}
  & 41 &  9, 32 \\
\cline{2-3}
  & 53 & 23, 30 \\
\cline{2-3}
  & 61 & 11, 50 \\
\cline{2-3}
  & 65 & 8,  18, 47, 57 \\
\cline{2-3}
  & 73 & 27, 46 \\
\cline{2-3}
  & 85 & 13, 38, 47, 72 \\
\cline{2-3}
  & 89 & 34, 55 \\
\cline{2-3}
  & 97 & 22, 75 \\
\cline{2-3}
  & 101 & 10, 91 \\
\cline{2-3}
  & 109 & 33, 76 \\
\cline{2-3}
  & 113 & 15, 98 \\
\cline{2-3}
  & 125 & 57, 68 \\
\cline{2-3}
  & 137 & 37, 100 \\
\cline{2-3}
  & 145 & 12, 17, 128, 133\\
\cline{2-3}
  & 149 & 44, 105 \\
\cline{2-3}
  & 157 & 28, 129 \\
\cline{2-3}
  & 169 & 70, 99 \\
\cline{2-3}
  & 173 & 80, 93 \\
\cline{2-3}
  & 181 & 19, 162 \\
\cline{2-3}
  & 185 & 43, 68, 117, 142 \\
\cline{2-3}
  & 193 & 81, 112 \\
\cline{2-3}
  & 197 & 14, 183 \\
\hline
5 & 11  & 3, 4, 5, 9 \\
\cline{2-3}
  & 31 & 2, 4, 8, 16 \\
\cline{2-3}
  & 41 & 10, 16, 18, 37 \\
\cline{2-3}
  & 61 & 9, 20, 34, 58 \\
\cline{2-3}
  & 71 & 5, 25, 54, 57 \\
\cline{2-3}
  & 101 & 36, 84, 87, 95 \\
\cline{2-3}
  & 121 & 3, 9, 27, 81 \\
\cline{2-3}
  & 131 & 53, 58, 61, 89 \\
\cline{2-3}
  & 151 & 8, 19, 59, 64 \\
\cline{2-3}
  & 181 & 42, 59, 125, 135 \\
\cline{2-3}
  & 191 & 39, 49, 109, 184 \\
\hline
\end{tabular}
\end{center}
\caption{Fulfillments under $3 \le \mathrm{ord}_P(\sigma) \le 5$ and $P \le 200$}
\label{hagi:tbl-fullfill-1}
\end{table}

\begin{table}[htb]
\begin{center}
\begin{tabular}{l|c|r}
\hline\hline
\multicolumn{1}{c|}{\bf $\mathrm{ord}_P (\sigma)$} & {\bf $ P $}& {\bf $\sigma$}\\
\hline
6 & 13 & 4, 10 \\
\cline{2-3}
  & 19 & 8, 12 \\
\cline{2-3}
  & 31 & 6, 26 \\
\cline{2-3}
  & 37 & 11, 27 \\
\cline{2-3}
  & 43 & 7, 37 \\
\cline{2-3}
  & 49 & 19, 31 \\
\cline{2-3}
  & 61 & 14, 48 \\
\cline{2-3}
  & 67 & 30, 38 \\
\cline{2-3}
  & 73 &  9, 65 \\
\cline{2-3}
  & 79 & 24, 56 \\
\cline{2-3}
  & 91 & 10, 17, 75, 82 \\
\cline{2-3}
  & 97 & 36, 62 \\
\cline{2-3}
  & 103 & 47, 57 \\
\cline{2-3}
  & 109 & 46, 64 \\
\cline{2-3}
  & 127 & 20, 108 \\
\cline{2-3}
  & 133 & 12, 31, 103, 122 \\
\cline{2-3}
  & 139 & 43, 97 \\
\cline{2-3}
  & 151 & 33, 119 \\
\cline{2-3}
  & 157 & 13, 145 \\
\cline{2-3}
  & 163 & 59, 105 \\
\cline{2-3}
  & 169 & 23, 147 \\
\cline{2-3}
  & 181 & 49, 133 \\
\cline{2-3}
  & 193 & 85, 109 \\
\cline{2-3}
  & 199 & 93, 107 \\
\hline
7 & 29  & 7, 16, 20, 23, 24, 25 \\
\cline{2-3}
  & 43  & 4, 11, 16, 21, 35, 41  \\
\cline{2-3}
  & 71  & 20, 30, 32, 37, 45, 48 \\
\cline{2-3}
  & 113 & 16, 28, 30, 49, 106, 109 \\
\cline{2-3}
  & 127 & 2, 4, 8, 16, 32, 64 \\
\cline{2-3}
  & 197 & 36, 104, 114, 164, 178, 191 \\
\hline
8 & 17 & 2, 8, 9, 16 \\
\cline{2-3}
  & 41 & 3, 14, 27, 38 \\
\cline{2-3}
  & 73 & 10, 22, 51, 63 \\
\cline{2-3}
  & 89 & 12, 37, 52, 77 \\
\cline{2-3}
  & 97 & 33, 47, 50, 64\\
\cline{2-3}
  & 113 & 18, 44, 69, 95 \\
\cline{2-3}
  & 137 & 10, 41, 96, 127 \\
\cline{2-3}
  & 193 & 9, 43, 150, 184 \\
\hline
9 & 19 & 4, 5, 6, 9, 16, 17 \\
\cline{2-3}
  & 37 & 7, 9, 12, 16, 33, 34 \\
\cline{2-3}
  & 73 & 2, 4, 16, 32, 37, 55 \\
\cline{2-3}
  & 109 & 16, 27, 38, 66, 75, 105 \\
\cline{2-3}
  & 127 & 22, 37, 52, 68, 99, 103 \\
\cline{2-3}
  & 163 & 38, 40, 53, 85, 133, 140\\
\cline{2-3}
  & 181 & 39, 43, 62, 65, 73, 80 \\
\cline{2-3}
  & 199 & 43, 58, 162, 175, 178, 180 \\
\hline
\end{tabular}
\end{center}
\caption{Fulfillments under $6 \le \mathrm{ord}_P (\sigma) \le 9$ and $P \le 200$}
\label{hagi:tbl-fullfill-2}
\end{table}

\begin{table}[htb]
\begin{center}
\begin{tabular}{l|c|r}
\hline\hline
\multicolumn{1}{c|}{\bf $\mathrm{ord}_P (\sigma)$} & {\bf $ P $}& {\bf $\sigma$}\\
\hline
10 & 31 & 15, 23, 27, 29 \\
\cline{2-3}
  & 41 & 4, 23, 25, 31 \\
\cline{2-3}
  & 61 & 3, 27, 41, 52 \\
\cline{2-3}
  & 71 & 14, 17, 46, 66 \\
\cline{2-3}
  & 101 & 6, 14, 17, 65 \\
\cline{2-3}
  & 121 & 40, 94, 112, 118 \\
\cline{2-3}
  & 131 & 42, 70, 73, 78 \\
\cline{2-3}
  & 151 & 87, 92, 132, 143 \\
\cline{2-3}
  & 181 & 46, 56, 122, 139 \\
\cline{2-3}
  & 191 & 7, 82, 142, 152\\
\hline
11 & 23 & 2, 3, 4, 6, 8, 9, 12, 13, 16, 18 \\
\cline{2-3}
   & 67 & 9, 14, 15, 22, 24, 25, 40, 59, 62, 64 \\
\cline{2-3}
   & 89 & 2, 4, 8, 16, 32, 39, 45, 64, 67, 78 \\
\cline{2-3}
   & 199 & 18, 61, 62, 63, 103, 114, 121, 125, 139, 188 \\
\hline
12 & 37 & 8, 14, 23, 29 \\
\cline{2-3}
   & 61 & 21, 29, 32, 40 \\
\cline{2-3}
   & 73 & 3, 24, 49, 70 \\
\cline{2-3}
   & 97 & 6, 16, 81, 91 \\
\cline{2-3}
   & 109 & 8, 41, 68, 101 \\
\cline{2-3}
   & 157 & 22, 50, 107, 135 \\
\cline{2-3}
  & 169 & 19, 80, 89, 150 \\
\cline{2-3}
  & 181 & 7, 26, 155, 174 \\
\cline{2-3}
  & 193 & 49, 63, 130, 144 \\
\hline
13 & 53 & 10, 13, 15, 16, 24, 28, 36, 42, 44, 46, 47, 49 \\
\cline{2-3}
   & 79 & 8, 10, 18, 21, 22, 38, 46, 52, 62, 64, 65, 67 \\
\cline{2-3}
   & 131 & 39, 45, 52, 60, 62, 63, 80, 84, 99, 107, 112, 113 \\
\cline{2-3}
   & 157 & 14, 16, 39, 46, 67, 75, 93, 99, 101, 130, 153 \\
\hline
14 & 29 & 4, 5, 6, 9, 13, 22 \\
\cline{2-3}
   & 43 & 2, 8, 22, 27, 32, 39 \\
\cline{2-3}
   & 71 & 23, 26, 34, 39, 41, 51 \\
\cline{2-3}
   & 113 & 4, 7, 64, 83, 85, 97 \\
\cline{2-3}
   & 127 & 63, 95, 111, 119, 123, 125 \\
\cline{2-3}
   & 197 & 6, 19, 33, 83, 93, 161 \\
\hline
15 & 31 & 7, 9, 10, 14, 18, 19, 20, 28 \\
\cline{2-3}
   & 61 & 12, 15, 16, 22, 25, 42, 56, 57 \\
\cline{2-3}
   & 151 & 2, 4, 16, 38, 76, 85, 105, 128\\
\cline{2-3}
   & 181 & 5, 25, 27, 29, 82, 114, 117, 145\\
\hline
\end{tabular}
\end{center}
\caption{Fulfillments under $10 \le \mathrm{ord}_P (\sigma) \le 15$ and $P \le 200$}
\label{hagi:tbl-fullfill-3}
\end{table}

\begin{table}[htb]
\begin{center}
\begin{tabular}{l|c|r}
\hline\hline
\multicolumn{1}{c|}{\bf $\mathrm{ord}_P (\sigma)$} & {\bf $ P $}& {\bf $\sigma$}\\
\hline
16 & 97 & 8, 12, 18, 27, 70, 79, 85, 89 \\
\cline{2-3}
  & 113 & 35, 40, 42, 48, 65, 71, 73, 78 \\
\cline{2-3}
  & 193 & 3, 27, 50, 64, 129, 143, 166, 190\\
\hline
17 & 103 & 8, 9, 13, 14, 23, 30, 34, 61,\\
   &     & 64, 66, 72, 76, 69, 81, 93, 100 \\
\cline{2-3}
   & 137 & 16, 34, 38, 50, 56, 59, 60, 72, 73, 74,\\
   &     & 88, 115, 119, 122, 123, 133 \\
\hline
18 & 37 & 3, 4, 21, 25, 28, 30 \\
\cline{2-3}
   & 73 & 18, 36, 41, 57, 69, 71 \\
\cline{2-3}
   & 109 & 4, 34, 43, 71, 82, 93 \\
\cline{2-3}
   & 127 & 24, 28, 59, 75, 90, 105 \\
\cline{2-3}
   & 163 & 23, 30, 78, 110, 123, 125 \\
\cline{2-3}
   & 181 & 101, 108, 116, 119, 138, 142 \\
\cline{2-3}
  & 199 & 19, 21, 24, 37, 141, 156 \\
\hline
19 & 191 & 5, 6, 25, 30, 32, 36, 52, 69, 107,\\
   &     & 121, 125, 136, 150, 153, 154, 160, 177, 180\\
\hline
20 & 41 & 2, 5, 8, 20, 21, 33, 36, 39 \\
\cline{2-3}
   & 61 & 8, 23, 24, 28, 33, 37, 38, 53 \\
\cline{2-3}
   & 101 & 32, 39, 41, 44, 57, 60, 62, 69 \\
\cline{2-3}
   & 181 & 22, 31, 35, 74, 107, 146, 150, 159 \\
\hline
\end{tabular}
\end{center}
\caption{Fulfillments under $16 \le \mathrm{ord}_P (\sigma) \le 20$ and $P \le 200$}
\label{hagi:tbl-fullfill-4}
\end{table}

We call a triple $(P, \sigma, \tau)$ a \textbf{perfume (PERfect FUlfillMEnt)} if the triple satisfy the following conditions:
\begin{itemize}
\item $\sigma $ is a fulfillment to $P$,
\item $\tau$ is coprime to $P$,
\item $\tau \neq \sigma, \sigma^2, \dots, \sigma^{\mathrm{ord}(\sigma)} $.
\end{itemize}

\begin{proposition}\label{hagi:ppp}
Let $P, \sigma, \tau$ be positive integers such that $(P, \sigma, \tau)$ is a perfume.
Then both of $( \sigma^{-a} - \sigma^{-b})$ and $\tau (\sigma^{-a} - \sigma^{-b})$ are coprime to $P$ for $0 \le a, b < \mathrm{ord}_{P} (\sigma)$ and $a \neq b$.
In other words, both of $( \sigma^{-a} - \sigma^{-b})$ and $\tau (\sigma^{-a} - \sigma^{-b})$ are elements of $\mathbb{Z}_P^* $.
In particular, there exists inverses of $( \sigma^{-a} - \sigma^{-b})$ and $\tau (\sigma^{-a} - \sigma^{-b})$.
\end{proposition}
\begin{proof}
First, we prove $(\sigma^{-a} - \sigma^{-b})$ is coprime to $P$.
Put $c$ as the maximal value of $\{ a , b \}$.
Then $(\sigma^{-a} - \sigma^{-b}) = \sigma^{-c} (\sigma^{c-a} - \sigma^{c-b})$.
Note that one of $\sigma^{c-a}$ and $\sigma^{c-b}$ is 1.
From the definition of fulfillment, $\sigma^{-c}$ and $(\sigma^{c-a} - \sigma^{c-b})$ are coprime to $P$.
Hence $(\sigma^{-a} - \sigma^{-b})$ is coprime to $P$.

Now $\tau$ is coprime to $P$.
Then $\tau (\sigma^{-a} - \sigma^{-b})$ is coprime to $P$ if and only if $(\sigma^{-a} - \sigma^{-b})$ is coprime to $P$.
\end{proof}

\subsection{Perfume and Four-cycle}
In the authors' opinion, the following theorem is the main contribution of this paper.
Remark that there is no other systematic construction for ingredient codes satisfy (G) and (T) simultaneously.

\begin{theorem}\label{hagi:main3}
Let $P$ be an integer $\ge 2$.
Let $(P, \sigma, \tau)$ be a perfume and $L$ the twice of the order of $\sigma$ as an element of a group $\mathbb{Z}_{P}^{*}$, i.e. $L=2 \mathrm{ord}(\sigma)$.

Put 
\[ c_{j,l} := 
\left\{
  \begin{array}{cc}
  \sigma^{-j + l}     &  0 \le l < L/2  \\
  \tau \sigma^{-j + l}     &  L/2 \le l < L,  \\
  \end{array}
\right.
\]
\[ d_{k,l} := 
\left\{
  \begin{array}{cc}
  - \tau \sigma^{k - l}     &  0 \le l < L/2  \\
  - \sigma^{k-l}     &  L/2 \le l < L  \\
  \end{array}
\right.
\]
and put model matrices
$\mathcal{H}_{C} = [ c_{j,l} ]_{0 \le j <J, 0 \le l < L}$, $\mathcal{H}_{D}=[ d_{k,l} ]_{0 \le k <K, 0 \le l < L},$
where $L = \mathrm{ord}( \sigma )$, $1 \le J, K \le L/2$.
Then associated QC-LDPC codes $(C, H_C)$ and $(D, H_D)$ satisfy (G) and (T).
In particular, the associated quantum code is a four-cycle code.
\end{theorem}
\begin{proof}[Proof]
Put tires $T_A$ and $T_B$ by:
\[
T_{A} =
\left(
  \begin{array}{cccc}
    1   & \sigma   & \dots   & \sigma^{L/2 -1}   \\
    \sigma^{L/2 -1}   & 1   &    & \sigma^{L/2 -2}   \\
    \vdots   &    & \ddots   & \vdots   \\
    \sigma   & \sigma^2   & \dots   & 1  \\
  \end{array}
\right),
\]
\[
T_{B} = 
\left(
  \begin{array}{cccc}
    \tau   &  \tau \sigma   & \dots   & \tau \sigma^{L/2 -1}   \\
    \tau \sigma^{L/2 -1}   & \tau   &    & \tau \sigma^{L/2 -2}   \\
    \vdots   &    & \ddots   & \vdots   \\
    \tau \sigma   & \tau \sigma^2   & \dots   & \tau  \\
  \end{array}
\right).
\]
Then $\mathcal{H}_C = [ T_A T_B ]$ and $\mathcal{H}_D = [ -T_B^{\mathrm{T}} -T_A^{\mathrm{T}} ]$.
Therefore we can apply Proposition \ref{hagi:prop:two_rs_mats} to $C$ and $D$.
Hence (T) holds.

Next, we prove that (G) holds for $C$.
Let $c_j$ be the $j$th row of $\mathcal{H}_C$.
Fix $0 \le a < b < J$ and denote the first $L/2$ entries of $c_{a} - c_{b}$ by $X$, and the remaining $L/2$ entries by $Y$.
Then we have 
$X = (\sigma^{-a} - \sigma^{-b}) \times (1, \sigma, \sigma^{2}, \dots, \sigma^{L/2 -1}) $ and
$ Y = (\sigma^{-a} - \sigma^{-b})\tau \times (1, \sigma, \sigma^2, \dots, \sigma^{L/2-1}) $,
where $[]$ is with notation in Proposition \ref{hagi:quotient}.
In other words, $X = [ \sigma^{-a} - \sigma^{-b} ]$ and $Y= [ \tau ( \sigma^{-a} - \sigma^{-b} )]$ as sets.

By Proposition \ref{hagi:prop:free}, the condition that $c_{a}-c_{b}$ be multiplicity free is equivalent to $[\sigma^{-a} - \sigma^{-b}] \neq [\tau(\sigma^{-a} - \sigma^{-b})]$.
It is obvious that $[\sigma^{-a} - \sigma^{-b}] \neq [\tau(\sigma^{-a} - \sigma^{-b})]$ if and only if $[1] \neq [\tau]$.
By the choice of $\tau$, $[1] \neq [\tau]$ holds.
\end{proof}

\begin{example}\label{hagi:expl1}
Let $P=7$, ($\mathbb{Z}_{P}^{*} = \{ 1, 2, 3, 4, 5, 6 \}$) and choose $\sigma = 2$ and $\tau = 3$.
Then $L/2 = \mathrm{ord}(2) = 3$ and we put $J=K=L/2$ ($=3$).
Then the parity-check matrices are obtained from Theorem \ref{hagi:main3}:
\[
\mathcal{H}_{C} =
\left(
  \begin{array}{cccccc}
    1   & 2   & 4   & 3   & 6   & 5   \\
    4   & 1   & 2   & 5   & 3   & 6   \\
    2   & 4   & 1   & 6   & 5   & 3   \\
  \end{array}
\right)
\]
\[
=
\left(
  \begin{array}{cccccc}
    1   & 2   & 2^2   & 3   & 3*2   & 3*2^2   \\
    2^2   & 1   & 2   & 3*2^2   & 3   & 3*2   \\
    2   & 2^2   & 1   & 3*2   & 3*2^2   & 3   \\
  \end{array}
\right),
\]
\[
\mathcal{H}_{D} = 
\left(
  \begin{array}{cccccc}
    4   & 2   & 1   & 6   & 3   & 5   \\
    1   & 4   & 2   & 5   & 6   & 3   \\
    2   & 1   & 4   & 3   & 5   & 6   \\
  \end{array}
\right)
\]
\[
=
\left(
  \begin{array}{cccccc}
    -3   & -5   & -6   & -1   & -4   & -2   \\
    -6   & -3   & -5   & -2   & -1   & -4   \\
    -5   & -6   & -3   & -4   & -2   & -1   \\
  \end{array}
\right).
\]
\end{example}

\subsection{Existence and Construction for Regular Ingredient Codes}
From here, we prove that there exists a large variety of regular LDPC ingredient codes.
Formally, we state the following:
\begin{theorem}\label{hagi:main1}
For any even $L > 0$ and any $1 \le J, K \le L/2$, there exist integers $P$ such that
\begin{itemize}
\item $(C , H_C) $ is a $(J, L)$-regular QC-LDPC code,
\item $(D , H_D) $ is a $(K, L)$-regular QC-LDPC code,
\item the size of circulant matrices is $P$,
\item $(C, H_C)$ and $(D, H_D)$ satisfy (G) and (T) simultaneously.
\end{itemize}
\end{theorem}
\begin{proof}[Proof of Theorem \ref{hagi:main1}]
We show the existence of a perfume $(P, \sigma, \tau)$ such that $\mathrm{ord}_P (\sigma) = L/2$.
It implies that there exists ingredient codes $C', D'$ such that they are $(L/2, L)$-regular QC-LDPC codes with circulant matrix size $P$.
Denote their model matrices by $\mathcal{H}_{C}'$ and $\mathcal{H}_{D}'$.
Put another model matrix $\mathcal{H}_C$ by deleting first $L/2 -J$ rows from $\mathcal{H}_{C}'$.
Then $\mathcal{H}_C$ is a matrix of size $J \times L$.
Therefore we obtain a $(J, L)$-regular QC-LDPC code $(C, H_C)$.
By a similar way, we obtain a $K \times L$ matrix $\mathcal{H}_D$ and a $(K, L)$-regular QC-LDPC code $(D, H_D)$.
Since $C'$ and $D'$ satisfy (G) and (T) simultaneously, $(C, H_C)$ and $(D, H_D)$ also satisfy (G) and (T) simultaneously.

Let us show the existence of a perfume $(P, \sigma, \tau)$.
In a case $L = 2$, $(3, 1, 2)$ is a perfume.
From here, we assume $L \ge 4$.
By Dirichlet's Theorem \ref{hagi:gauss} certify the existence of a prime $P = 1 + (L/2) n$ by some positive integer $n$.
Since $P$ is a prime, $\mathbb{Z}_{P}^{*}$ is a cyclic group of order $(L/2) n$, by Proposition \ref{hagi:cyclic}.
Let $z$ be a generator of $\mathbb{Z}_{P}^{*}$.
Put $\sigma := z^n$, then we have $\mathrm{ord}(\sigma) = L/2$.
Since $P$ is a prime, any element in $\mathbb{Z}_{P} \setminus \{0\} = \mathbb{Z}_P^*$ is coprime to $P$.
In particular, $\sigma$ is a fulfillment to $P$.

A set $\{ \sigma, \sigma^2 , \dots, \sigma^{\mathrm{ord}(\sigma)} \}$ consists of $L/2$ elements.
We have $\# ( \mathbb{Z}_P^{*} \setminus \{ \sigma, \sigma^2 , \dots, \sigma^{\mathrm{ord}(\sigma)} \} ) = (n-1) L/2$:
in particular, $\mathbb{Z}_P^{*} \setminus \{ \sigma, \sigma^2 , \dots, \sigma^{\mathrm{ord}(\sigma)} \}$ is not an empty set.
Thus we can pick $\tau$ up from $\mathbb{Z}_{P}^{*} \setminus \{1, \sigma, \sigma^2, \dots \}$.
Therefore, we obtain a perfume $(P, \sigma, \tau)$.
\end{proof}

\begin{example}
For a perfume $(P, \sigma, \tau) = (101, 95, 2)$ with $\mathrm{ord}_{P}(\sigma) = 5$,
we obtain two $(5, 10)$-regular QC-LDPC codes $(C, H_C)$ and $(D, H_D)$ if we choose $J= K = 5$.

Their model matrices are $\mathcal{H}_C$ and $\mathcal{H}_D$:
\[
\left(
  \begin{array}{cccccccccc}
1 & 95 & 36 & 87 & 84 & 2 & 89 & 72 & 73 & 67 \\
84 & 1 & 95 & 36 & 87 & 67 & 2 & 89 & 72 & 73 \\
87 & 84 & 1 & 95 & 36 & 73 & 67 & 2 & 89 & 72 \\
36 & 87 & 84 & 1 & 95 & 72 & 73 & 67 & 2 & 89 \\
95 & 36 & 87 & 84 & 1 & 89 & 72 & 73 & 67 & 2 \\
  \end{array}
\right)
\]
and
\[
\left(
  \begin{array}{cccccccccc}
99 & 34 & 28 & 29 & 12 & 100 & 17 & 14 & 65 & 6 \\
12 & 99 & 34 & 28 & 29 & 6 & 100 & 17 & 14 & 65 \\
29 & 12 & 99 & 34 & 28 & 65 & 6 & 100 & 17 & 14 \\
28 & 29 & 12 & 99 & 34 & 14 & 65 & 6 & 100 & 17 \\
34 & 28 & 29 & 12 & 99 & 17 & 14 & 65 & 6 & 100 \\
  \end{array}
\right).
\]

Let $M_C = (m_{c0}, m_{c1}, \dots) $ and $M_D= (m_{d0}, m_{d1}, \dots) $ be a binary vector with length $\mathrm{ord}_{P}(\sigma)$, which is $5$ here.
We call $M_C$ and $M_D$ \textbf{mask vectors}.
We delete the $i$th row from $\mathcal{H}_C$ (resp. $\mathcal{H}_D $) if $m_{ci} = 0$ (resp. $m_{di}=0$).
For $M_{C} = (1, 1, 1, 0, 1)$ and $M_{D} = (0,1,0,1,1)$,
we obtain
\[
\left(
  \begin{array}{cccccccccc}
1 & 95 & 36 & 87 & 84 & 2 & 89 & 72 & 73 & 67 \\
84 & 1 & 95 & 36 & 87 & 67 & 2 & 89 & 72 & 73 \\
87 & 84 & 1 & 95 & 36 & 73 & 67 & 2 & 89 & 72 \\
95 & 36 & 87 & 84 & 1 & 89 & 72 & 73 & 67 & 2 \\
  \end{array}
\right)
\]
and
\[
\left(
  \begin{array}{cccccccccc}
12 & 99 & 34 & 28 & 29 & 6 & 100 & 17 & 14 & 65 \\
28 & 29 & 12 & 99 & 34 & 14 & 65 & 6 & 100 & 17 \\
34 & 28 & 29 & 12 & 99 & 17 & 14 & 65 & 6 & 100 \\
  \end{array}
\right).
\]
Then we obtain $( \mathrm{wt}(M_C), L)$-regular QC-LDPC code and $( \mathrm{wt}(M_D), L)$-regular QC-LDPC code which satisfy (G) and (T) simultaneously, where $\mathrm{wt}(x)$ is the Hamming weight of $x$.
By using mask vectors, it is easy to characterize model matrices for ingredient codes.
\end{example}

\section{Analysis on Our Proposal --Variations, Optimality and Word Error Rate--}

\subsection{Any Code Rates are Possible by Our Construction}
Recall that the quantum code rate $R$ of a CSS code with ingredient codes $C$ and $D$ is defined by the following:
$$ R := \frac{\dim C - \dim D^{\perp}}{n}, $$
where $\dim C$ and $\dim D^{\perp}$ are the dimensions of $C$ and $D^{\perp}$ as linear spaces over $\mathbb{F}_2$ respectively and $n$ is the code length of $C$ and $D$.

By Theorem \ref{hagi:main1}, we obtain a $(J, L)$-regular QC-LDPC code $(C, H_C)$ and a $(K, L)$-regular QC-LDPC code $(D, H_D)$ for any even $L$ and $1 \le J,K \le L/2$.
In general, the dimensions $\dim C$ of QC-LDPC code $C$ is almost the same as $LP - (JP-J+1)$.
This implies that the quantum code rate $R$ obtained by Theorem \ref{hagi:main1} is
$$R \simeq 1 - (JP+KP-J-K+2)/LP \simeq \frac{L - (J+K)}{L}.$$
For applying a Sum-Product decoding which is known as a standard decoding algorithm for LDPC codes, it is required to satisfy $J, K \ge 2$, i.e. $J+K \ge 4$.
Therefore, approximate possible quantum code rates are
$$ 0, \frac{1}{L}, \frac{2}{L}, \dots, \frac{L-4}{L}.$$

We show that it is possible to achieve approximately any quantum code rate $k/n$ for any integers $0 \le k < n$.
Choose a positive even integer $2m$ such that $2m(n-k) \ge 4$.
By Theorem \ref{hagi:main1}, we can construct ingredient codes $(C, H_C)$ and $(D, H_D)$ such that $(C, H_C)$ is a $(J, L)$-regular QC-LDPC code, $(D, H_D)$ is a $(K, L)$-regular QC-LDPC code, $L=2mn$, and  $J=K=m(n-k)$.

Then we have 
$$ R \simeq \frac{L-(J+K)}{L} = \frac{ 2mk }{ 2mn} = \frac{k}{n}.$$
Note that  $J+K \ge 4$ follows, by the choice of $2m$.

Therefore our construction provides any code rate approximately.

\subsection{Performance of Error-Correction}

Generally speaking, a protocol of quantum communication with error-correction consists of the following steps:
\begin{itemize}
\item \textit{Set up} Prepare a message state,
\item \textit{Encoding} Encode the message state to quantum code state,
\item \textit{Sending} The quantum code state is sent to a receiver,
\item \textit{Measurement} The receiver performs measurement for the received state and obtains outcomes, which are called a syndrome,
\item \textit{Error-Correction} According to the syndrome, the receiver performs recovery operation for the measured state and obtains an estimated state.
\end{itemize}
By measurement process, outcomes are two $\{1, -1\}$-sequences $O_C$ and $O_D$ which are associated with $H_C$ and $H_D$ respectively.
The receiver replaces $1$ to $0$ and $-1$ to $1$ in $O_C$ (resp. $O_D$) and obtains a syndrome $s_C$ (resp. $s_D$).
In quantum LDPC cases, to perform recovery operation, we use a syndrome sum-product decoding $\mathrm{SynDec}$ (see details in \cite{hagi:isit2010}).
$\mathrm{SynDec}$ receives syndromes $s_C$ and $s_D$, it outputs binary sequences $e_C$ and $e_D$.
Then we obtain a recovery operation $X^{e_D} Z^{e_C}$, where $X$ (resp. $Z$) is a bit-flip operation (resp. a phase operation),  $X^{e} := X^{e1} \otimes X^{e2} \otimes \dots \otimes X^{n}$ for $e = (e1, e2, \dots, en)$ and $Z^{e}$ is defined in a similar manner.

\begin{figure*}[htbp]
  \begin{center}
    \includegraphics[keepaspectratio=true,height=70mm]{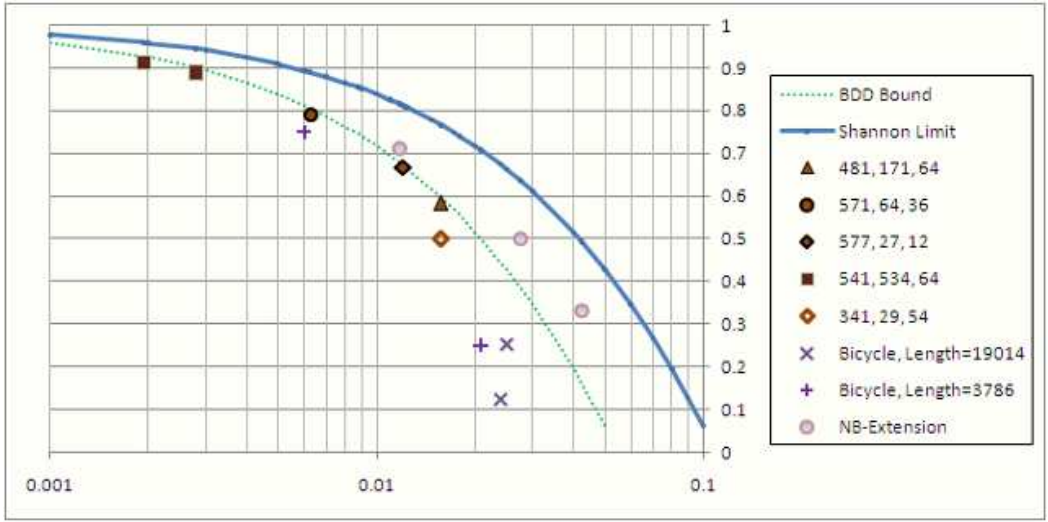}
  \end{center}
  \caption{Quantum Code Rate and Cross Over Probability $p$ at Failure Fidelity Rate $10^{-4}$ }%
  \label{hagi:fig:four_cycle_experience}
\end{figure*}

We performed computer experiences for evaluating error-correcting performance of our codes over quantum channel with sum-product decoding, which is a standard decoding method for LDPC codes.
For the experiences, we adopt a Pauli channel with bit-flip error probability $p -p^2$, phase error probability $p -p^2$, and combined error probability $p^2$ where $p$ is a positive real number and called \textbf{cross over probability} here.
This Pauli channel is called \textbf{two independent binary symmetric channels} in \cite{hagi:mackay}.
We chose the maximal iteration number as $128$.
We have to note that if we choose bigger iteration number, we obtain better error-correcting performance.
On the other hand, it takes more computational costs.
In quantum communication, a fidelity $f$ is one of criteria to evaluate of communication reliability.
The value $f$ is closely related to performance of quantum dense coding, quantum teleportation and the security of quantum key distribution.
The higher fidelity $f$ is, the better performance is.
In Fig \ref{hagi:fig:four_cycle_experience}, the point is plotted with the quantum rate as vertical axis and with the cross over probability as horizontal axis.
If a point is on $(x, y)$, it implies the following:
the code achieves the average fidelity $10^{-4}$ for cross over probability $x$ and its quantum code rate is $y$.

The dotted line shows one of famous benchmarks, called a bounded distance decoding (BDD) bound, or called a VG bound.
As it is stated in \cite{hagi:mackay}, the line is widely believed to be the maximal rate at which a bounded distance decoder can communicate even if a code has a possible largest minimum distance.
The BDD bound is generally obtained by $1- h(2p)$ for classical cross over probability $p$ over classical binary symmetric channel, where $h()$ is a binary entropy function.
For quantum communication, the bound is modified to $1-2 h(2p)$ for cross over probability over two independent symmetric channels if the classical rates of ingredient codes are the same to each other.
In a similar manner, Shannon limit is modified to $1-2 h(p)$.

Figure \ref{hagi:fig:four_cycle_experience} shows that four-cycle codes almost achieve the BDD (VG) bound at the quantum  code rates $0.9, 0.8, 0.7, 0.6$.
We may clarify some of model matrices of these four-cycle codes.
For example:
\begin{itemize}
\item perfume $(571, 64, 36)$ and its mask vectors are $M_C = (1, 0, 0, 0, 0, 0, 0, 0, 1, 1, 0, 0, 0, 1, 0, 0, 0, 0, 0)$ and $M_D = (0, 0, 0, 0, 0, 1, 0, 0, 0, 1, 1, 0, 0, 0, 0, 0, 0, 0, 1)$. The related quantum code rate is about 0.78975.
\item perfume $(577, 27, 12)$ and its mask vectors are $M_C = (1, 0, 1, 1, 0, 0, 0, 0, 1, 0, 0, 0)$ and $M_D = (0, 0, 0, 1, 0, 0, 0, 0, 1, 1, 0, 1)$. The related quantum code rate is about 0.6671.
\end{itemize}

Dots associated with NB-extension show outstanding error-correcting performance.
The extension method is proposed by Kasai et. al \cite{hagi:kasai}.
They extend our four cycle codes from a point of view of non-binary extension.
The possible code rate $R_Q$ of NB-extension is only $1 - 2/L$ for some positive integer $L$, i.e. $R_Q = 0, 1/3, 1/2, 3/5, 2/3, 5/7, \dots$.
Therefore there are different advantages among NB-extension codes and our four cycle code, e.g. various code rate, error-correcting performance.

\subsection{Optimality on The Code Length}
It is known that a necessary condition to have ``girth $> 4$ in the Tanner graph representation of a quasi-cyclic LDPC code $(C, H_C)$ with $J$ row-blocks and $L$ column-blocks'' is ``$P \ge L+1$'' for even $L$ and $J \ge 3$, where $P$ is the size of the circulant matrices. (See \cite{hagi:ieice,hagi:Fossorier}.)
This bound is known as a tight bound for many $L$ and $J=3$.

We would like to investigate a similar bound for a quantum case.
Let $(C, H_C)$ and $(D, H_D)$ be $(J, L)$-regular and $(K, L)$-regular QC-LDPC codes respectively such that $J, K \ge 3$, $L$ is an even integer and they satisfy (T) and (G) simultaneously.
A question is ``how large $P$ is required?'', where $P$ is the size of circulant matrices for $(C, H_C)$ and $(D, H_D)$.
Since we require additional conditions, it is natural to expect $P$ becomes larger than classical case.
However we obtain the following:
\begin{theorem}
``$P \ge L+1$'' is also tight bound for CSS codes with ingredient QC-LDPC codes if $L+1$ is a prime.
\end{theorem}
\begin{proof}
Put $P := L+1$.
There exists a generator $g$ of $\mathbb{Z}_{P}^{*}$ i.e. $\mathbb{Z}_{P}^{*} = \{ 1, g, \dots, g^{P-1} \}.$
Then $(P, g^2, g)$ is a perfume.
We have $\mathrm{ord}(g^2) = L/2$.
By applying Theorem \ref{hagi:main3},
we obtain ingredient QC-LDPC codes such that the size of circulant matrices is $L+1$.
\end{proof}
It surprises us that the bound on the length for a QC-LDPC code is still tight for quantum quasi-cyclic LDPC codes under additional condition (T).

\subsection{Difference Matrix Theory}
Let $\mathbb{Z}_L$ be an cyclic group of order $L$.
A cyclic difference matrix based on $\mathbb{Z}_L$, denoted $(L, J;1)$-CDM, is a $J \times L$ matrix $( p_{j,l} ),  p_{j,l} \in \mathbb{Z}_L$, such that for each $1 \le r < s \le J$, the differences $a_{r, l} - a_{s, l}$, $1 \le l \le L$ comprise all the elements of $\mathbb{Z}_{L}$.

Difference matrices \cite{ge,crc} have been studied in the construction of orthogonal arrays \cite{6}, the construction of authentication codes \cite{29}, software testing \cite{11,12}, data compression \cite{27}, general Steiner triple systems related to constant weight codes \cite{33}.

For $P=L$ and a matrix $C$, a $C$ is a $(L, J;1)$-CDM if and only if the difference of any rows of $C$ is multiplicity free, i.e. (G) holds.
Thus the condition (G) is regarded as a natural generalization of the definition of CDM.
On the other hand, the condition (T) is also regarded as another extension of the definition of CDM.

The difference matrix theory has been widely studied and there are so many constructions of them have been proposed.
It is expected to find the quantum code construction method from the difference matrix theory or its related areas.

\section{Conclusion}
We propose a class of quantum LDPC codes and call it four-cycle code.
Properties of the codes are proved by a characterization as Theorem \ref{hagi:iff:twisted} and Proposition \ref{hagi:prop:free}.
We show that four-cycle codes are generalizations of bicycle codes, its ingredient codes are regular LDPC codes, our proposed code satisfy a sort of optimality, and the performances almost achieve a BDD bound.
Designing weight distribution is an interesting problem from a point of view of not only mathematical structure but also error-correcting performance.
In fact, the study of weight distribution is arisen from evaluation of error-correcting performance for LDPC codes \cite{hagi:richardson}.
In \cite{hagi:kasai}, Kasai used our result for applying non-binary extension.
The key idea of their paper is to use $(2, L)$-regular LDPC codes.

The remaining cases for regular weight distribution are
\begin{itemize}
\item row weights are different,
\item row weights are odd,
\item row weights are the same and even, and one of column weights exceeds the half of the row weight.
\end{itemize}
These cases are open problems.
There may exist a method of design theory, algebraic combinatorics and so on.

\section*{Acknowledgments}
This research was partially supported by Grants-in-Aid for Young Scientists (B), 18700017, 2006.
We thank Dr. Yuichiro Fujiwara for valuable comments on the bridge between quasi-cyclic LDPC code theory and difference matrix theory, Dr. Kenta Kasai for discussion on non-binary extension, and S. Aly for e-mail communication.

\end{document}